\PassOptionsToPackage{table}{xcolor}

\documentclass[acmsmall, anonymous=false]{acmart}

\usepackage{ltablex}
\usepackage{graphicx}
\usepackage{titlesec}
\usepackage{xcolor}
\usepackage{natbib}
\usepackage[ruled]{algorithm2e} 
\usepackage{etoolbox}
\usepackage[all]{nowidow}
\usepackage{subcaption}
\usepackage[utf8x]{inputenc}
\usepackage{textcomp}
\usepackage[export]{adjustbox}
\usepackage{listings}
\usepackage{wrapfig}
\usepackage{mwe}
\usepackage{caption}
\usepackage{subcaption}
\usepackage{tablefootnote}
\usepackage{array,tabularx,ragged2e,booktabs}
\usepackage[width=.9\textwidth]{caption}
\usepackage{multirow}

\AtBeginDocument{%
  \providecommand\BibTeX{{%
    \normalfont B\kern-0.5em{\scshape i\kern-0.25em b}\kern-0.8em\TeX}}}

\definecolor{pink}{HTML}{EFEFEF}

\setcopyright{rightsretained}
\acmJournal{PACMHCI}
\acmYear{2021} \acmVolume{5} \acmNumber{CSCW2} \acmArticle{368} \acmMonth{10} \acmPrice{}\acmDOI{10.1145/3479512}

\begin{document}

\title{A Framework of Severity for Harmful Content Online}


\author{Morgan Klaus Scheuerman}
\affiliation{
  \institution{University of Colorado Boulder}
  \department{Department of Information Science}
  \state{CO}
  \country{USA}
  }
\email{morgan.scheuerman@colorado.edu}

\author{Jialun Aaron Jiang}
\affiliation{
  \institution{University of Colorado Boulder}
  \department{Department of Information Science}
  \state{CO}
  \country{USA}
  }
\email{aaron.jiang@colorado.edu}

\author{Casey Fiesler}
\affiliation{
  \institution{University of Colorado Boulder}
  \department{Department of Information Science}
  \city{Boulder}
  \state{CO}
  \country{USA}
  }
\email{casey.fiesler@colorado.edu}

\author{Jed R. Brubaker}
\affiliation{
  \institution{University of Colorado Boulder}
  \department{Department of Information Science}
  \city{Boulder}
  \state{CO}
  \country{USA}
  }
\email{jed.brubaker@colorado.edu}

\renewcommand{\shortauthors}{Morgan Klaus Scheuerman et al.}


\begin{abstract}
The proliferation of harmful content on online social media platforms has necessitated empirical understandings of experiences of harm online and the development of practices for harm mitigation. Both understandings of harm and approaches to mitigating that harm, often through content moderation, have implicitly embedded frameworks of prioritization---what forms of harm should be researched, how policy on harmful content should be implemented, and how harmful content should be moderated. To aid efforts of better understanding the variety of online harms, how they relate to one another, and how to prioritize harms relevant to research, policy, and practice, we present a theoretical framework of severity for harmful online content. By employing a grounded theory approach, we developed a framework of severity based on interviews and card-sorting activities conducted with 52 participants over the course of ten months. Through our analysis, we identified four Types of Harm (physical, emotional, relational, and financial) and eight Dimensions along which the severity of harm can be understood (perspectives, intent, agency, experience, scale, urgency, vulnerability, sphere). We describe how our framework can be applied to both research and policy settings towards deeper understandings of specific forms of harm (e.g., harassment) and prioritization frameworks when implementing policies encompassing many forms of harm. 
\end{abstract}

\ccsdesc[500]{Human-centered computing~Empirical studies in HCI}
\ccsdesc[500]{Human-centered computing~HCI theory, concepts and models}

\keywords{Safety, harm, framework, online communities, content moderation, social networks, severity.}


\maketitle

\begin{center}
\setlength\fboxsep{.5cm}
\fbox{%
	\parbox{0.8\linewidth}{%
		\textbf{CONTENT WARNING:} This paper contains sensitive and potentially triggering content, including discussions of death, sexual assault, emotional trauma, self-injury, eating disorders, and animal abuse.%
	}%
}
\end{center}

\section{Introduction}
In 2014, a blog post that attacked and made false claims about game developer Zo\"{e} Quinn started a series of controversies and events later famously known as Gamergate \cite{Aghazadeh2018,Braithwaite2016,Massanari2020}. Prior research, as well as journalistic writings, broadly characterized Gamergate as a harassment campaign that induced vast emotional harm. However, the real consequences and harm caused by Gamergate were much more complex than ``harassment,'' and the people it impacted also extended beyond Zo\"{e} Quinn: multiple women who came to Quinn's defense received prolonged rape and death threats \cite{Wingfield2014}. Among them, Anita Sarkeesian, a feminist media critic, had to cancel her appearance at Utah State University, which received threats of ``massacre'' against all attendees including students and staff \cite{Robertson2014}. However, in stark contrast, Nathan Grayson, the reporter alleged to have exchanged favorable game reviews for a sexual relationship with Quinn, only received accusations of journalistic misconduct. He did not suffer continued harassment, let alone death or rape threats. Taken together, these incidents clearly show that the harm in Gamergate was complex: it involved multiple kinds of harm, in different levels of severity for different people.

As demonstrated by the example above, alongside the benefits of online communities has been the  proliferation of harmful content and interactions, ranging from hate speech (e.g., \cite{Halder2014}) to violent imagery (e.g., \cite{Livingstone2014}) to destructive misinformation (e.g., \cite{Fernandez2018}). Social computing researchers have often focused on some forms of harmful content and interactions which are particularly pervasive and often experienced by end-users, such as interpersonal harms like harassment (e.g., \cite{Pater2016,Blackwell2019,Nova2018}) and bullying (e.g., \cite{Kwak2015,Singh2017,Ashktorab2016a}). Prior work has also uncovered deeply contextual experiences of interpersonal harm for specific groups, like women \cite{Vitak2017} and LGBTQ people \cite{Scheuerman2018a}. HCI has also benefited from in-depth research on those publishing and engaging with objectionably harmful content, like those looking to recover from eating disorders \cite{Chancellor2016} and self-harm \cite{Kruzan2020}, and the often negative experiences participants have when that content is moderated. Other harms are often hidden from the public. Professional content moderators, in particular, work to take down illegal and dangerous content quickly, to avoid legal repercussions and so that the public is not exposed to it \cite{Roberts,Koebler2018}. 

Scholarship across multiple disciplines has demonstrated various kinds of harm that can occur online, many of which have not been addressed specifically within HCI and social computing scholarship (e.g., child exploitation, terrorist coordination). However, it is important to understand not just what harms are, but how they occur in relation to one another. 

Gamergate, an example of a complex set of different harms that affect different people at different levels, shows that simply recognizing that a behavior \emph{is} harmful is not enough---it is also important to understand \emph{how} harmful the behavior is. Further, when considering multiple kinds of harm, we contend that it is crucial to understand severity of harms compared to others. After all, it clearly would be wrong to equate accusations of journalistic misconduct with death threats. 
When harms are co-occurring, particularly on large-scale social media platforms, understanding the relationships between them is crucial to deeper analyses of experiences of harm, prioritization practices for moderators, and more informed engagement with the perceptions of end-users on harm mitigation practices.

To understand the severity of a variety of experiences of online harms, we conducted a study to determine what makes a harm more or less severe. Given the complexity of harm, we felt it was crucial to capture a wide range of perspectives; this includes both experts who deal with harm and consider severity as part of their jobs (e.g., in content moderation) and everyday people who may come into contact with harmful content online. We asked participants to assess a range of harms present on---and often prohibited from---online platforms. Through our analysis, we identified how severity shaped perceptions of harmful online content. 

We employed a grounded theory approach to develop a framework based on empirical data collected over a period of ten months. We conducted both semi-structured interviews and card-sorting activities with 52 participants. We recruited 40 ``expert'' participants based on their specific expertise in a variety of professional areas related to harm (e.g., content moderation policy, mental health, etc.). The remaining 12 participants were ``general population'' social media users, whose broad experiences on social media were applied to both expand on and triangulate the perspectives of expert participants. 

Through our analysis, we identified four Types of Harm (physical, emotional, relational, and financial) and eight Dimensions along which the severity of harm can be understood (perspectives, intent, agency, experience, scale, urgency, vulnerability, sphere). We present our findings in the form of a framework for researchers and practitioners, so that they can determine the \emph{severity} of online harms---whether some harms are worse than others and what makes them worse. Our framework provides a tool to address the severity of differing harms as a set of complex, contextual, and overlapping factors.

We situate our framework by first describing the related work on social computing on online harm and content moderation. We then describe prioritization frameworks in domains outside of social computing and HCI. Next, we present our theoretical framework of severity, including both the Types of Harm and contextual Dimensions that shape severity. Through our discussion, we describe the relational and compounding nature of both Types of Harm and the Dimensions that make up the framework. We demonstrate the utility of the severity framework to researchers and practitioners, particularly those focused on online harm and content moderation, on both specified forms of harm (e.g., harassment) and on relationships between multiple harms (e.g., self-injury, hate speech, and child pornography).

\section{Related Work}

We situate our work within two areas of social computing. First, we review prior research on experiences of harm in online spaces, focusing on what definitions and perspectives on harm prior work has contributed to social computing. Second, we engage with prior work on content moderation, particularly practices employed to mitigate online harms and the necessity for frameworks that guide more effective mitigation. 

\subsection{Safety from Harm: Expansive Definitions of Online Harm in HCI}

While online safety and experiences of harm have become a large focus of CSCW and HCI research, these concepts tend to be defined contextually, often allowing technology users to define for themselves what it means to be safe \cite{Redmiles2019}. However, taken together, safety typically means protection from harm, which can be perceived as a risk, injury, or some undesirable outcome (e.g., \cite{Pater2016,Scheuerman2018a}). As such, the concept of ``safety'' from ``harm'' in a digital space has been adopted for numerous perspectives. 

One common perspective considers safety from \textit{technical} risk: for example, concerns about account integrity, including phishing scams \cite{Kumaraguru2007} and hacking \cite{Shakarian2016} attempts by bad actors. Hacking, in particular, has been studied for both its implications for account security \cite{Zangerle2014} and also as a method of harassment \cite{Matthews2017a}. There have also been emerging concerns around the intentional use of social media for spreading disinformation \cite{Hjorth2019} and attempting to upend political elections \cite{Dutt2019}. These technical perspectives on safety position harm as potentially adversely impacting individuals' digital possessions or accounts \cite{Fernandez2018}, financial well-being \cite{Amin2016}, and even reputation \cite{Lawson2015}. As such, policies surrounding account and election integrity have begun to be adopted by large social media platforms. For example, Twitter has adopted a policy stating that its users ``may not use Twitter's services for the purpose of manipulating or interfering in elections'' \cite{Twitter}. 

Within social computing research, the largest focus of safety has been on targeted interpersonal harm, such as bullying, hate speech, and harassment (e.g. \cite{Blackwell2018,Poland2016}). In particular, researchers have examined experiences of interpersonal harm on different groups, such as children \cite{Schrock2009,Ringland2015}, women \cite{Vitak2017,Nova2018}, and LGBTQ people \cite{Scheuerman2018a, Fernandez2019}. Pater et al. illuminated the lack of clear definitions of interpersonal harm by platform policies, which often conflate ``hate,'' ``bullying,'' ``harassment,'' and ``abuse'' \cite{Pater2016}. Yet, scholars focused on hate speech and harassment see them as distinct. While harassment is the targeted aggression towards an individual or group of individuals \cite{Ballard2017,Singh2017,Blackwell2019}, hate speech is identity-specific expressions of hatred or violence towards people of a certain race, ethnicity, gender, or sexuality \cite{Chaudhry2015a,Chatzakou2017,Phadke2018}.

Content-based harm has been another focus of social computing scholarship. Content-based harm refers to the harm caused to individuals who view undesirable content on a social media platform. For example, content that is triggering for victims of trauma \cite{Tynes2019} or reinforces the reality of racial violence for people of color \cite{Scheuerman2018a, To2020} would result in content-based harm. As such, not all digital harm has an interpersonal component, as one might expect with harassment or bullying. One well researched area is around the use of social media for self-harm. Pater et al. defined digital self-harm as ``online communication and activity that leads to, supports, or exacerbates, non-suicidal yet intentional harm or impairment of an individual’s physical wellbeing'' \cite{Pater2016}. Digital self-harm can be linked to posting or seeking out self-harming content, such as images of cutting \cite{Pater2016,Andalibi2017b} or content promoting anorexia \cite{Haas2011,Chancellor2016}. Given the potentially harmful nature of this content, many platforms have adopted policies against content that promotes self-harm---although the difference between therapeutic and promotional content can be difficult to discern \cite{Chancellor2016}.

Within social computing, there are a few existing frameworks that define relationships of harm, but none have considered the severity of that harm. Scheuerman et al. describe harm as consisting of six overlapping facets: outsider, insider, targeted, incidental, individual, and collective \cite{Scheuerman2018a}. While Scheuerman et al.'s framework can be used to classify the many different forms of harm described above, it remains unclear how to operationalize the \textit{moderation} of different forms of harms. Further, it is difficult to discern when harm is most impactful and when. In this work, we build on existing research on the wide range of digital harms that can occur on social media. In the next section, we discuss the current work in HCI and technology studies on online content moderation. Content moderation is an important area of research and practice for understanding current harm mitigation practices on social media platforms, and how a severity framework would aid necessary prioritization approaches.

\subsection{Content Moderation as Harm Mitigation: Opportunities for a Severity-Based Approach}

Beyond experiences of safety---and lack thereof---on online communities, social computing scholars have specifically focused on content moderation to better understand methods of mitigating online harm. Content moderation refers to the practice of controlling unwanted content in online spaces, whether that content is viewed as simply irrelevant (e.g., to an online forum with a specific topic), obscene, or illegal \cite{Fiesler2017}. Content moderation is currently the predominant method for addressing harm on online platforms. 
 Moderation often relies on two sorts of guiding documents for managing harm: a small-community's rules, which are determined by the administrators of subcommunities on larger platforms (e.g., a subreddit on Reddit) or a platform's terms of use, which determine what forms of content and interactions are allowed on the platform at all (e.g., Facebook's Community Standards). Content is often moderated in a variety of ways, from warnings issued about a specific piece of content without its removal, to removal of the content itself, to the banning of the user from a subcommunity or platform entirely \cite{Seering2019,Gillespie2018}. 

Moderation can occur in three ways, often in tandem: volunteer moderation, paid commercial moderation, and automatic algorithmic moderation \cite{Gillespie2018}. Research on small community moderation practices has been incredibly rich, given the accessibility researchers have to volunteer moderators in comparison to commercial moderators. Volunteer moderators are often users of a platform who have committed their time, labor, and expertise to supporting the moderation practices of a small subcommunity (e.g., subreddits, Facebook groups). These volunteer moderators have extensive administrative power over their own communities, such as setting rules, removing content, and banning people \cite{Seering2019}. They are most focused on mitigating interpersonal harms within their subcommunities, such as hate speech or harassment (e.g., \cite{Wohn2019,Jiang2019}). Volunteer moderators' scope is relegated only to subcommunities, and they cannot mitigate harm on a platform-level. Further, they do not generally assess content that is relegated to commercial moderators, like child pornography, beheading videos, and animal abuse.

On the other hand, paid commercial moderators work to remove content at the platform level; they are sometimes hired in-house (e.g., they work for Twitter), or they work as contractors with no direct employment relationship with the platform (e.g., they work for a company like Cognizant and are contracted by Twitter). Many platforms are incentivized to moderate: not only to meet legal and policy requirements, but also to avoid losing users subject to malicious behaviors, to protect their corporate image, to placate advertisers who do not want to be associated with ``sketchy'' online communities, and to honor their own institutional ethics \cite{Gillespie2018,Klonick2018}. Gillespie posits that professional content moderators are critical resources---if not \emph{the} critical resource---that social media platforms offer their users, due to their crucial role in mitigating harm \cite{Gillespie2018}. Much of the content viewed by commercial moderators is traumatic \cite{Roberts}. Roberts, through ethnographic work on commercial content moderation, uncovered the emotional toll that moderating the Internet's darkest content---such as violent videos---takes on commercial moderators \cite{Roberts}, highlighting the capacity of harm the forms of content posted online carries.
Currently, the vast majority of research on commercial content moderation is illuminating end-user perspectives on moderation practices---generally, uncovering the dissatisfaction with end users with platform moderation (e.g., \cite{Feuston2020,Blackwell2017a, Nurik2019,MyersWest2018}). However, recent work has uncovered that users perceive both commercially-moderated and volunteer-moderated communities as similarly toxic, suggesting current human moderation practices are failing to reduce negative perceptions of harm online \cite{Cook2021}.

As online communities grew into sizes that human moderation could not reasonably handle, automated moderation provided a hopeful solution for moderation at scale \cite{Chandrasekharan2019}. Automated moderation (``automoderation'') is the process of automatically flagging or removing violating content through trained algorithms, rather than relying on human moderators, and has been used in both commercial and volunteer moderation. While automated moderation can ideally work quickly and effectively at removing harmful content, it suffers from a lack of nuanced or cultural understanding that often results in failures in the ``gray areas'' of harm mitigation, such as hate speech \cite{Jhaver2019}. This not only results in some hate speech remaining on the platform, but users having negative perceptions about what platforms characterize as harmful \cite{Blackwell2017a,Nurik2019}.
Ruckenstein and Turunen argue that the benefits of machines is that they cannot be emotionally harmed by the inhumane content regularly removed from platforms, while human moderators are currently subject to that harm. However, machine learning methods are generally supervised to be more effective, thus still requiring human labor in annotating large amounts of data \cite{Ruckenstein2020}. Automoderation therefore suffers from similar labor concerns, including repetitive exposure to inhumane and traumatic content, as human moderation approaches \cite{Dang2018,Hammer2019,Vincent2019}.


While moderation is currently the solution to mitigating harmful content and interactions online, ideally keeping most of that content away from general users, there are difficulties in enacting appropriate harm mitigation at scale and in ways all parties agree with. Given issues of prioritization and labor have been persistent in online moderation, how can we better understand the relationships of different harms to better triage? To improve moderation practices, one possible approach is to consider the severity of different harms. Prioritization frameworks focused on the severity of harm could ease human moderator capacity and focus automoderation training efforts, so that the worst content could be handled first. Our framework on the severity of online harms offers a taxonomy that can be employed and scaled for a variety of content moderation approaches, and it also provides new lenses through which to examine moderator practices and experiences. 

\section{Assessing the Severity of Harm in Other Domains}
\label{domains}

Many other domains also utilize prioritization frameworks to streamline processes, allocate resources, and assess impact. Such perspectives offer valuable insight to social computing and content moderation, which necessitates decision-making around what forms of harm to research, how to build policies for a wide variety of harms, and how content moderation labor is prioritized. In the final section of our related work, we discuss prioritization frameworks in other domains and how they might inform a framework of online severity. To guide our work, we did an initial literature review of how other disciplines and professions approach severity. While our analysis is not exhaustive, we briefly present frameworks of severity in other domains here for two reasons. First, they provide alternative frameworks from which to contextualize our work. Second, these domains guided our interviews with some of our expert participants, who worked as practitioners in these domains, as we delved more deeply into how their work shaped their perspectives on severity. Specifically, we present a brief overview of three practice-based domains and their approaches to severity: (1) law, (2) law enforcement, and (3) mental health.

When considering severity in law, particularly with the judicial system, we can see a \emph{punitive approach} to severity. Here, sentencing correlates to the amount of harm a person has caused. A familiar example might be how murder is assessed in the United States. Though varying by state, a common differentiation is murder by ``degrees,'' where first-degree murder describes intentionality in the planning of the crime, while second-degree murder typically describes unplanned killing or death resulting from reckless disregard for human life. Alongside these degrees are differing sentencing limitations, written into law with increasing severity of punishment depending on the degree of the crime. When mapping the punitive model to content moderation, moderation actions---such as warning or banning---are commonly associated with the ``degree'' of the violation. 

Local law enforcement, meanwhile, often triages using a \emph{time-based approach} when dispatching offers. Due to the wide range of incoming calls, 911 operators, police dispatchers, and officers have to prioritize tasks based on a ``severity'' criteria that prioritizes danger to human beings. Calls move through a pipeline based on the information provided by the caller, starting first at the 911 operator, then queued to a police dispatcher, and then making its way to an officer for response. When prioritizing calls, operators and police follow strict risk hierarchies, foregoing a ``first come, first serve'' to focus on the most severe cases first. ``Priority codes'' define how quickly a police offer should respond to a call, where the number ``1'' usually represents the highest priority (e.g., violent felonies in progress, shootings, fellow police officer down) (e.g., \cite{Gilgor,CityofEugenePoliceD2009,Police2016}). Priority 1, or P-1, calls require an immediate response—while the response time required becomes less immediate as priorities numbers increase (e.g., P-3). Lower priority calls often make up the majority of calls received, necessitating a perspective on severity that prioritizes fewer calls with higher risk \cite{Asher}. If attempting to implement a moderation framework like that of law enforcement, content deemed urgent to address would be prioritized. 

Finally, mental health professionals assess severity through an \emph{persistence approach} focused on assessing the needs of an individual based on the duration of their symptoms and the physical threat they pose to themselves and others. Their methods for assessment are highly consistent, documented in the International Classification of Diseases (ICD-10) or, in the United States, the Diagnostic and Statistical Manual of Mental Disorders (DSM). The ICD and DSM code the severity of mental illness, where some mental illnesses are considered more severe than others. The timespan of mental illness and impact on individual wellbeing largely determine severity. Determining the severity of a mental illness involves assessing the duration and the debilitation of the symptoms an individual is experiencing. Long-term symptoms that make it difficult for the individual to operate are considered the most severe, while short-term symptoms that do not greatly impact daily life are considered the least severe. Similarly, the severity of self-harming behaviors is also often measured by well-documented scales that take into account factors like the duration of behavior, the frequency of self-harm acts, and the physical danger those acts pose. A mental health moderation framework would assess the number of times an individual engages in harmful behavior, and whether that behavior presents a physical threat to themselves or others.

Beyond these three, many other domains provide both practice-based and theory-based prioritization frameworks as well. For example, we might examine homelessness prevention to understand prioritization frameworks constrained by lack of resources that force allocation decisions. We might also look to feminist perspectives on restorative-justice for frameworks that prioritize survivors of harm before punitive measurements. 

Regardless of the source, these frameworks often represent an idealized outline for researchers and practitioners to follow, but often break down in practice; many may not be as clearly followed as guidelines or law dictate, resulting in a degradation of public trust and the efficacy of domain-based services. For example, failures of law enforcement to adequately protect citizens has reached a cultural crisis in the United States (failures of law enforcement) (e.g., \cite{Speri,Safronova2019,Peeples2020}). Similarly, punitive measures taken in the United States judicial system have been heavily criticized for mishandling sexual assault cases (e.g., \cite{Aycock2019,Alexander2020}). However, even with their limitations, frameworks like these can provide a common ground around which practitioners can organize and prioritize their efforts. Likewise, they can also provide a publicly available baseline that can promote accountability. Towards this end, we asked: what makes content harmful, and how can we assess the severity of that harm? 

\section{Methods}

To understand the relationships between the wide variety of harms that occur online, we conducted a rich interview (with qualitative card-sorting) study with two groups of participants. First, we interviewed experts who work in content moderation, platform policy, and occupations focused on harm assessment and mitigation (e.g., law enforcement, mental health research). Expert participants involved in moderation and platform policy had deep contextual knowledge about how platforms currently handle and think through harm mitigation practices. Those who worked in other areas of harm mitigation, like mental health researchers, offered unique perspectives on how other domains outside of online platforms think through harmfulness and the severity of harm. Second, we interviewed general population participants who use social media regularly. General population users offer a broader insight on what it is like to actually interact with the platform interfaces where harm occurs, and are the ``end users'' harm mitigation practices are meant to protect. Our interviews and card-sorting activities focused on how a wide range of people from differing backgrounds assessed the relationships between different forms of harm, particularly what makes some harm particularly ``bad.''

Through these interviews, we discovered numerous overlapping themes of harm that indicated how participants weighted, prioritized, and assessed harmfulness. We aimed to understand \textit{what} made certain harms more severe, rather than \textit{which} specific harms were considered most severe. Our analysis showed how participants assessed the \emph{severity} of harm, what aspects of a harmful interaction or harmful content increased the impact of that harm. We embraced a grounded-theory inspired approach to develop a framework for the severity of the harm of online content \cite{Charmaz2014}. Our process involved first gathering broad, open-ended information and then iteratively analyzing the data before deductively validating our themes. Our data consisted of semi-structured interviews with participants, including a card sorting activity. 

\begin{table}[h!]
\centering
\caption{\small Due to the scale of methods employed in this study, the table above simplifies each step into simple phases. Participant recruitment (as described in Section \ref{participants}) covered two participant groups: experts (divided into 3 Sub-Groups) and general population. Data collection (as described in Section \ref{design}) occurred in two phases within each of these participant groups: inductive and deductive. Data analysis (as described in Section \ref{data-analysis}) occurred iteratively as we collected data from all 4 participant groups.}
\label{table:phases}  
\def\tabularxcolumn#1{m{#1}}
\begin{tabularx}{\linewidth}{ccX}

        \multicolumn{3}{c}{\cellcolor[HTML]{EFEFEF}\textbf{Method Phases}} 
        \\ 
        \textit{\textbf{Step}} 
        & \multicolumn{1}{c}         
        {\textit{\textbf{Phases}}}
        & \multicolumn{1}{c}         
        {\textit{\textbf{Details}}}
        \\
        \cmidrule(lr){1-1}
        \cmidrule(lr){2-2}
        \cmidrule(lr){3-3}

        \rowcolor{pink}  & & \textbf{Phase One}: Interviews with 3 expert sub-groups. 
        \\
         \rowcolor{pink} \multirow{-2}{*}{Participant Recruitment}& \multirow{-2}{*}{Two}& \textbf{Phase Two}: Interviews with gen pop. 
        \\
        \multirow{3}{*}{Data Collection} & \multirow{3}{*}{Two} & For each participant group, we inductively analyzed data during interviews, then switched to deductive interviewing aimed at confirming emerging themes.
        \\
        & & \textbf{Phase One}: Inductive interviewing within subgroups.
        \\
        & & \textbf{Phase Two}: Deductive interviewing within subgroups.
        \\
         
        \rowcolor{pink}&  & \textbf{Phase One}: Iteratively analyze and confirm data from Sub-Group 1. 
        \\
          \rowcolor{pink}& & \textbf{Phase Two}: Iteratively analyze and confirm data from Sub-Group 2 as compared to themes from Sub-Group 1. 
        \\
          \rowcolor{pink}& & \textbf{Phase Three}: Iteratively analyze and confirm data from Sub-Group 3 as compared to themes from Sub-Groups 1 \& 2. Conduct a comparative analysis of themes from Sub-Groups 1 \& 3 in the context of industry expertise. 
        \\
          \rowcolor{pink} \multirow{-8}{*}{Data Analysis}& \multirow{-8}{*}{Four} & \textbf{Phase Four}: Iteratively analyze and confirm data from Gen Pop. Conduct a Comparative analysis of themes between all Expert Sub-Groups and Gen Pop. \\
        

\end{tabularx}
  
\end{table}

In this section, we describe (1) participant recruitment; (2) the design of the card sorting activity, including the creation of the categories of content we assessed; (3) the design of the interview protocols; and (4) our data analysis process. 

\subsection{Participant Recruitment}
\label{participants}

\begin{table}[htbp!]
\centering
\caption{A table of all participants, including their role in the study (expert or general population (``gen pop'')), their gender, their high-level expertise area, the context they work in, and the region they are situated. All industry experts worked in the context of content moderation, including but not limited to: the creation of community guidelines, research on content impacts, management of employees or projects, and processes relevant to moderation. Individuals were bucketed into high level expertise areas to protect their identities. Asterisks denote participants who acted as "community partners," providing member checks to the researchers during the research process. Participants in academia or non-profits were assigned more granular expertise areas---in particular, the areas for which they were recruited for.}
\label{table:participants}    
\small
\def\tabularxcolumn#1{m{#1}}

\begin{tabularx}{\textwidth}{XXXXXX}
        \textit{\textbf{Number}} 
        & \multicolumn{1}{c}         
        {\textit{\textbf{Type}}}
        & \multicolumn{1}{c}         
        {\textit{\textbf{Gender}}}
        & \multicolumn{1}{c}
        {\textit{\textbf{Expertise}}} 
        & \multicolumn{1}{c}
        {\textit{\textbf{Context}}} 
        & \multicolumn{1}{c}
        {\textit{\textbf{Region}}} 
        \\
        \cmidrule(lr){1-1}
        \cmidrule(lr){2-2}
        \cmidrule(lr){3-3}
        \cmidrule(lr){4-4}
        \cmidrule(lr){5-5}
        \cmidrule(lr){6-6}

        \rowcolor[HTML]{EFEFEF} 
        E1 & Expert & Woman & Policy & Industry &North America
        \\
        E2 & Expert & Man & Research & Industry & North America
        \\
        \rowcolor[HTML]{EFEFEF} 
        E3 & Expert & Woman & Mental Health & Industry & North America
        \\
        E4 & Expert & Man & Mental Health & Industry & North America
        \\
        \rowcolor[HTML]{EFEFEF} 
        E5 & Expert & Woman & Management & Industry & Western Europe
        \\
        E6 & Expert & Woman & Policy & Industry & North America
        \\
        \rowcolor[HTML]{EFEFEF} 
        E7 & Expert & Man & Management & Industry & Western Europe
        \\
        E8 & Expert & Man & Policy & Industry &  North America
        \\
        \rowcolor[HTML]{EFEFEF} 
        E9* & Expert & Woman & Policy & Industry & North America
        \\
        E10* & Expert & Woman & Policy & Industry & North America
        \\
        \rowcolor[HTML]{EFEFEF} 
        E11 & Expert & Woman & Policy & Industry & North America
        \\
        E12 & Expert & Woman & Policy & Industry & North America
        \\
        \rowcolor[HTML]{EFEFEF} 
        E13 & Expert & Man & Engineering & Industry & Western Europe
        \\
        E14 & Expert & Woman & Policy & Industry &  North America
        \\
        \rowcolor[HTML]{EFEFEF} 
        E15 & Expert & Man & Research & Industry & North America
        \\
        E16 & Expert & Woman & Moderation & Academia & North America
        \\
        \rowcolor[HTML]{EFEFEF} 
        E17 & Expert & Woman & Mental Health & Academia & South Asia
        \\
        E18 & Expert & Woman & Feminism & Academia & North America
        \\
        \rowcolor[HTML]{EFEFEF} 
        E19 & Expert & Man & Law & Academia & North America
        \\
        E20 & Expert & Woman & Mental Health & Academia & North America
        \\
        \rowcolor[HTML]{EFEFEF} 
        E21 & Expert & Man & Moderation & Academia & North America
        \\
        E22 & Expert & Woman & Feminism & Non-Profit & North America
        \\
        \rowcolor[HTML]{EFEFEF} 
        E23 & Expert & Man & Law & Law Enforcement & North America
        \\
        E24 & Expert & Woman & Moderation & Academia & North America
        \\
        \rowcolor[HTML]{EFEFEF} 
        E25 & Expert & Man & Policy & Industry & South Asia
        \\
        E26 & Expert & Woman & Policy & Industry & South Asia
        \\
        \rowcolor[HTML]{EFEFEF} 
        E27 & Expert & Woman & Policy & Industry & South Asia
        \\
        E28 & Expert & Woman & Policy & Industry & South Asia
        \\
        \rowcolor[HTML]{EFEFEF} 
        E29 & Expert & Man & Policy & Industry & South Asia
        \\
        E30 & Expert & Woman & Policy & Industry & South Asia
        \\
        \rowcolor[HTML]{EFEFEF} 
        E31 & Expert & Woman & Policy & Industry & South Asia
        \\
        E32 & Expert & Man & Policy & Industry & South Asia
        \\
        \rowcolor[HTML]{EFEFEF} 
        E33 & Expert & Man & Policy & Industry & South Asia
        \\
        E34 & Expert & Man & Policy & Industry & South Asia
        \\
        \rowcolor[HTML]{EFEFEF} 
        E35 & Expert & Man & Policy & Industry & South Asia
        \\
        E36 & Expert & Woman & Policy & Industry & South Asia
        \\
        \rowcolor[HTML]{EFEFEF} 
        E37 & Expert & Woman & Policy & Industry & South Asia
        \\
        E38 & Expert & Man & Policy & Industry & South Asia
        \\
        \rowcolor[HTML]{EFEFEF} 
        E39 & Expert & Woman & Policy & Industry & South Asia
        \\
        E40 & Expert & Woman & Policy & Industry & South Asia
        \\

\end{tabularx}

\end{table}

\begin{table}[htbp!]
\centering
\small
\def\tabularxcolumn#1{m{#1}}

\begin{tabularx}{\textwidth}{XXXXXX}
        \textit{\textbf{Number}} 
        & \multicolumn{1}{c}         
        {\textit{\textbf{Type}}}
        & \multicolumn{1}{c}         
        {\textit{\textbf{Gender}}}
        & \multicolumn{1}{c}
        {\textit{\textbf{Expertise}}} 
        & \multicolumn{1}{c}
        {\textit{\textbf{Context}}} 
        & \multicolumn{1}{c}
        {\textit{\textbf{Region}}} 
        \\
        \cmidrule(lr){1-1}
        \cmidrule(lr){2-2}
        \cmidrule(lr){3-3}
        \cmidrule(lr){4-4}
        \cmidrule(lr){5-5}
        \cmidrule(lr){6-6}

        \rowcolor[HTML]{EFEFEF} 
        G1 & Gen Pop & Woman & N/A  & N/A  & North America
        \\
        G2 & Gen Pop & Woman & N/A  &  N/A & North America
        \\ 
        \rowcolor[HTML]{EFEFEF} 
        G3 & Gen Pop & (Trans) Woman & N/A  & N/A & North America
        \\
        G4 & Gen Pop & Woman & N/A  &  N/A& North America
        \\
        \rowcolor[HTML]{EFEFEF} 
        G5 & Gen Pop & Woman & N/A  &  N/A& North America
        \\
        G6 & Gen Pop & Woman &  N/A & N/A & North America
        \\
        \rowcolor[HTML]{EFEFEF} 
        G7 & Gen Pop & Man & N/A  & N/A & North America
        \\
        G8 & Gen Pop & Man &  N/A & N/A & North America
        \\
        \rowcolor[HTML]{EFEFEF} 
        G9 & Gen Pop & Man & N/A  & N/A & North America
        \\
        G10 & Gen Pop & Man & N/A  & N/A & North America
        \\
        \rowcolor[HTML]{EFEFEF} 
        G11 & Gen Pop & Man & N/A  & N/A & North America
        \\
        G12 & Gen Pop & Non-Binary &  N/A & N/A & North America
        \\

\end{tabularx}

\end{table}

We recruited a total of 52 participants over two phases of data collection. All interviews were conducted in English. Participants were located in North America, Western Europe, and South Asia. Participants included 29 women, 22 men, and 1 non-binary person. We recruited participants using three methods: direct recruitment via email, a recruitment survey on social media, and snowballing \cite{Corbin2014}. A number of participants spoke to us under the understanding that we would not reveal their specific positions. We instead report on what is most important for this work: their professional expertise. We bucketed participants into broad categories that do not reveal their individual job titles, but showcase their expertise on moderation, harm, and social media guidelines. The expertise area of "policy" indicates an expertise with community guidelines, whether through development, research, or moderation. In general, participants can be divided into four sub-groups: (1) industry experts from Company 1 (E1-E15); (2) non-industry experts (E16-E24); (3) industry experts from Company 2 (E25-E40); and (4) general population users (G1-G12). A full table of all participants, their demographics, and their expertise areas can be found in Table \ref{table:participants}.

During our first phase of data collection, we recruited 40 experts with diverse experience related to content moderation. Company names were anonymized as requested by the companies for researcher access and to also protect individual employees. Similarly, we indicate broader regions for each participant to avoid specific associations with companies and individuals. We recruited experts from Company 1 directly via email and through snowballing with community partners. We recruited experts from Company 2 by reaching out to the company to arrange interviews with employees. All of the expert participants interviewed during this phase are referenced in our findings using the letter `E' (e.g., E17). Overall, this phase of expert interviews occurred in three stages. First, we interviewed experts working at a large North American tech company (``Company 1'' henceforth, 15 participants). Then, we interviewed experts working in contexts outside of industry, primarily academia (9 participants). Third, we interviewed experts working at an international technology consulting company in South Asia, geographically where a large part of the content policy workforce in the social media space is located \cite{Roberts} (``Company 2'' henceforth, 16 participants).

The majority of our interviews were with professionals working in the content moderation and policy space as they have unique, specific expertise when considering how to handle different content online. These interviews included individuals who worked on content moderation and policy teams, whose roles ranged from policy making to research on harm to mental health support ($n=35$). 




We also wanted to ensure we included other perspectives outside of content moderation, harassment research, and platform policy. As such, we recruited 5 additional participants with expertise in the areas mentioned in section \ref{domains}. These participants included a lawyer ($n=1$); a law enforcement officer ($n=1$); a feminist scholar ($n=1$); and mental health researchers ($n=2$). Experts in law, feminism, and health provided specific perspectives in their domains. For example, our lawyer participant discussed local and regional law, while our feminist participant discussed severity from a survivor-centered stance. We recruited these participants through direct email solicitation. Potential participants were identified by the research team through personal contacts, snowball sampling, and searching websites like Google and the ACM for key terms (e.g., ``content moderation researchers''). 

During phase two of data collection, we recruited ``general population'' participants. Given that the general population is the intended user of social media platforms, we felt is was important to understand their perspectives on harm and severity as well. Further, general population users sit on the other side of the screen from industry experts, and can offer insight into viewing and experiencing harmful content. Therefore, the data of 6 general population participants contributed to the inductive analysis of severity. while the remaining data of 6 general population participants contributed to the deductive analysis. These participants were recruited using two methods. The first was through a Qualtrics Panel, a service provided by Qualtrics for recruiting participants. The second was through social media recruitment and snowballing. We posted a recruitment survey on social media platforms like Facebook, Twitter, and Reddit; we also asked contacts to share the survey. We targeted participants that were 18 and older, given the sensitive nature of the interviews. We indicate general population participants in our findings using the letter `G' (e.g., G6). 

\subsection{Interview Design}
\label{design}

Semi-structured interviews were aimed at understanding concepts of severity when relating harmful or violating content online. We chose to conduct semi-structured interviews to gather rich perspectives from each participant in relation to their own contextual experiences, either as ``domain experts'' or general social media users. Interviews awarded us the opportunity to ask clarifying questions, seek out examples, and investigate participants' experiences with violating content in their own lives \cite{Seidman1998}. 

We conducted interviews iteratively over the course of ten months. We conducted all interviews using video conferencing software. We audio recorded interviews with participant consent using the affordances embedded in the conferencing software. While we acknowledge that video conferencing services limited participants with slower connections or lack of access to laptop or desktop computers from participating in the study, we chose to use video conferencing so participants could share their screens while doing the card sorting activity. Screen sharing allowed us to follow along with the card sort, noting any shifting of the card sorts in real time, so that we could follow-up with questions. We also asked participants to ``think aloud'' while conducting card sorts \cite{Charters2003}.

\begin{figure*}[htb!] 
    \centering
    \textbf{Data Analysis and Theory Formation Process}\par\medskip
    \includegraphics[width=\textwidth]{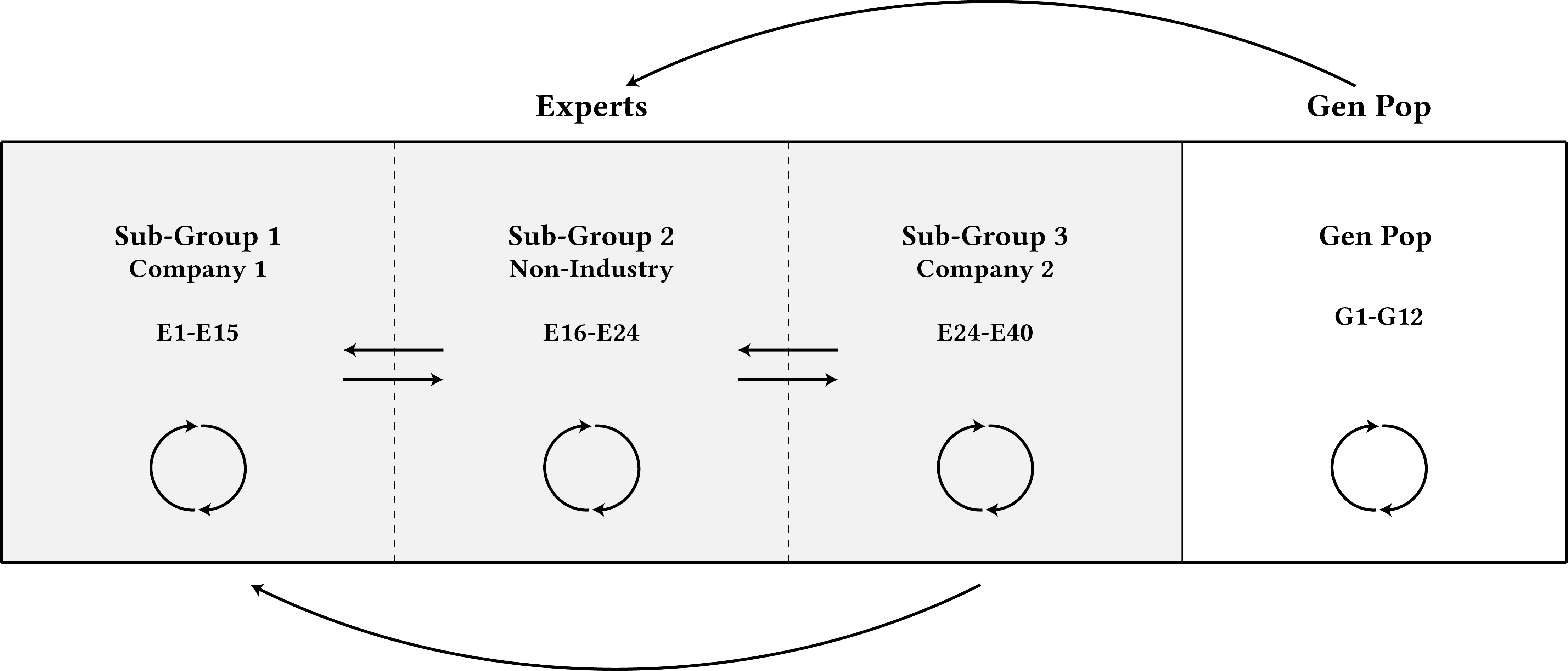}
    \caption{The figure above shows how we analyzed data within participant sub-groups and across participant sub-groups. Expert participant sub-groups are shown in gray, while general population participants are shown in white. The circular arrows represent the constant comparison of induction and deduction within each sub-group. The horizontal arrows between Sub-Group 1 \& Sub-Group 2 and Sub-Group 2 \& Sub-Group 3 represent the comparisons made between themes as they emerged between groups. We also compared Sub-Group 3 themes to Sub-Group 1, as they were different companies in different regions. Finally, we compared themes from Gen Pop to aggregated themes from all Experts to develop the finalized Framework of Severity.}
    \label{methods-fig}
\end{figure*}

Interviews were conducted in two phases for each sub-group of participants. How these phases co-occurred with analysis is shown in Figure \ref{methods-fig} in Section \ref{data-analysis}. The first phase was open-ended and exploratory. During this phase, we still regularly open coded and wrote memos about our findings \cite{Charmaz2014}. Interview questions were designed to first establish a general conception of severity in the eyes of each participant. For example, ``How do you define harm (online or offline)? How would you determine if one harm is worse than another?'' We then asked detailed questions of each card in the card sort. For example, ``Why did you place (that card) in that place? What makes it more severe than the above card? What makes it less severe than the above card?'' We iteratively coded these interviews to induce thematic conceptions of severity until we reached salience \cite{Charmaz2014}. 

The second phase was was aimed at confirming the thematic categories that arose during the first phase, and adjusting them with any additional information gleaned \cite{Charmaz2014}. These interviews were still focused on the card sorting activity, with a focus on deriving the most high-level reasoning behind participants' rankings of severity. At the end, we asked participants how they felt about the themes derived during the first phase of interviews. For example, ``Did you consider (theme) while ranking your cards?'' and ``Does the medium of content impact the severity of harm?'' Through our interviews with the general population, we confirmed the majority of the themes previously coded in our expert interview sub-groups. No new themes arose through our interviews with general population participants, but they offered unique perspectives on the themes we had already identified---including different perspectives on content-level rankings in the card sorts. We discuss these differences more in Section \ref{genpop-vs-expert}.

The phase process was iterative across the two interview phases. So for each sub-group, we iteratively coded and analyzed transcripts as we interviewed, moving into confirming themes as we reached saturation. Each additional sub-group of participants was contributed further to thematic saturation, with general population participants triangulating the themes observed in expert phases of interviews. A diagram of interview and analysis phases can be seen in Figure \ref{methods-fig}.

\subsection{Content Categories for Card Sorting}

Alongside interviews, we conducted a card sorting activity with participants. Card sorting is a usability technique often employed by user experience designers and information architects to understand how users would organize or rank categories of data or information \cite{Spencer2009}. We conducted both interviews and card sorting with each participant in the same session. The categories we used for card sorting are shown in Table \ref{table:cards}.

Card sorting allowed us the opportunity to understand severity using concrete examples of content already moderated on most social media platforms. We probed participants during interviews on their perspectives on severity in the context of their card-sorting decisions. We were able to inquire about their decisions to rank certain content as more or less severe than others, and also to clarify whether they believed the severity of content should be considered in rank-order or in ranked groups. In other words, we were able to clarify whether or not participants felt some content was equally severe and why they might feel that way. This approach informed both our deeper understanding of our card sorting data, contextualizing it beyond numerical averages for which to rank content, and the deductive phase of our data analysis, in which we applied qualitative concepts to a potential policy ranking of severity \cite{Charters2003}.

We used categories of content from Jiang et al., who conducted a content analysis of eleven social media platforms with the most monthly active users based on published statistics, which revealed 66 different policy violations across eleven different social media platforms \cite{JialunAaronJiang2020}. We conducted a secondary thematic grouping of the 66 violations that resulted in 20 more abstract content categories. Some platforms already do this. For example, Facebook, which seems to have the most extensive list of policies \cite{JialunAaronJiang2020}, breaks down what they call ``Violence and Criminal Behavior'' into five different categories of policy \cite{Facebook}. 

We do acknowledge that not all categories of potentially harmful content may be present in our final list. As our goal was not to rank granular content by severity, but rather to assess high-level concepts about what might make some policies more or less severe, we felt this simplification was appropriate. It also made it more manageable for participants to card sort and discuss. Further, not all content categories are actively enforced on social media websites---meaning, not all content is actually removed from websites if it is posted (e.g., child nudity). However, regardless of whether the content is deemed as removable by moderation teams, websites maintain policies on such content. 

We also found it ethically problematic to show participants---even those working as experts in the moderation space---examples of specific content (e.g., harassment posts, images of graphic violence). Beyond that, some content is illegal to possess (e.g., child pornography) and may be considered highly traumatic to view, as we did discover through our findings. Therefore, all card categories instead contain textual examples, as seen in Table \ref{table:cards}. The researchers also clarified any examples with participants as they went through the card sorting activity. We regularly referred to the specific examples to ensure participant understanding, as well. Therefore, we believe the limitations of not assessing actual moderated content examples was appropriate both methodologically and ethically.

\begin{table}[h!]
\centering
\caption{\small A table showing the content categories used in the card sorting. For each high-level category violation (e.g., ``Platform Abuse'') we also provide examples of that violation (e.g., spam, fake accounts). When conducting the card sorting activity and interviews with participants, each card contained examples of content violations.}
\label{table:cards} 
\rowcolors{2}{white}{pink}
\def\tabularxcolumn#1{m{#1}}
\begin{tabularx}{\linewidth}{lX}
        \multicolumn{2}{c}{\cellcolor[HTML]{EFEFEF}\textbf{Content Categories and Examples}} 
        \\ 
        \cmidrule(lr){1-2}
        \textit{\textbf{Category}} 
        & \multicolumn{1}{l}         
        {\textit{\textbf{Examples}}}
        \\
        \cmidrule(lr){1-1}
        \cmidrule(lr){2-2}
         
        Adult Sexual Exploitation & Sextortion\footnotemark[1], adult non-consensual sexual images 
        \\
        Platform Abuse & Spam, fake accounts
        \\ \footnotetext{Sextortion is a form of sexual exploitation that involves coercing or bribing a person for sexual favors, often by threatening to release intimate details or photos to the public, to the victims' workplace, or to the victims' loves ones.}
         
        Targeted Attacks on an Individual Person & Bullying, harassment \\
        
        Pornography & Pornographic images of adults, sexual texts
        \\
        Coordinated Attacks that Promote Harming People & Terrorist propaganda, hate groups, celebration of mass murder
        \\
         
        Violence & Credible threats of injury or murder
        \\
        Self-Injury and Eating Disorder Promotion & Images or posts of pro-anorexic content
        \\
         
        Graphic Violence Imagery & Images of gore, sadistically celebrating violent images
        \\
        Direct Harm to Children & Child pornography, child grooming
        \\
         
        Animal Abuse & Videos or images of animals being abused
        \\
        Sale of Firearms and Sex Work (``Prostitution''\footnotemark[2]) & Things that are illegal or highly regulated to sell online
        \\
         
        Hate Speech & Slurs, encouraging violence against a protected group
        \\
        Coordinating Scams and Political Attacks & Misinformation, fraud, voter suppression
        \\

        Hard Drug Sales & Advertisements for the selling of drugs like cocaine and heroin
        \\
        Informative Images of Graphic Violence & Disturbing journalistic images of war
        \\
       
       Suicide Promotion & Urging people to kill themselves, posting videos of suicide that promote suicide
       \\
        Regulated Goods & Selling marijuana, tobacco, or alcohol
        \\
        
        Child Nudity & Images parents post of their children in a bathtub
        \\
        Adult Non-Sexual Nudity and Digital Nudity & Images of nude adults, drawings of nude adults
        \\
         
        Coordinated Attacks that Directly Harm People & Human trafficking, mass murder coordination
    \end{tabularx}
    \vspace{-5mm}
\end{table}

\footnotetext[1]{Sextortion is a form of sexual exploitation that involves coercing or bribing a person for sexual favors, often by threatening to release intimate details or photos to the public, to the victims' workplace, or to the victims' loves ones.}
\footnotetext[2]{Many content policies group these together as things that are illegal to sell on their services. Many policies also use the terms ``prostitution'' and it was often more familiar to our participants, which is why we decided to keep the term on our cards. This category was determined with the help of our community partners, who practice policy in a North American tech corporation and are adopting North American perspectives on legality.}

Each participant was asked to rank the above categories by severity using the online tool, OptimalSort \cite{optimal}. Cards were randomized for each participant, mitigating ordering effects. We asked participants to sort from most severe (1) to least severe (20). Participants could either sort cards into rank order (1, 2, 3, ...) or in ranked groups (1, 1, 2, 3, 3...), in cases where they felt some content was ``equally severe.'' Each participant was asked to ``submit'' their card sort data through OptimalSort. We then exported the card sort data to a spreadsheet for each participant so that we could conduct data analysis. 

\subsection{Data Analysis}
\label{data-analysis}

We adopted constructivist grounded theory methods in conducting and analyzing our research \cite{Charmaz2014}. We open-coded the first phase of data collection in three iterations, using each iteration to further our knowledge of our data and inform our coding scheme. During the second, deductive phase of data collection, we explicitly asked participants questions prompted by our framework. Our aim was to increase our data coverage across the various themes while also vetting our analysis during the first phase. We also open-coded this phase of data collection, confirming previous codes and developing new ones.

Through each iteration, we coded each participant transcript in segments with lower level ideas found in the data. We then later conducted focused coding to group these data into higher-level categories. These categories reflected factors that participants regularly referenced or considered when determining severity across the forms of harm we discussed. Next, we derived two sets of themes by analyzing the relationship between these categories \cite{Saldana}. In this paper, we present these two sets as our severity framework. Our first set of themes describe the Types of Harm, abstracted collections of different forms of harmful content and interactions which may be analyzed through a lens of severity. The themes from the second set describe Dimensions, what shapes the severity of Types of Harm or more granular instances of harm.

Throughout this process we wrote and re-wrote memos describing our categories and themes, discussing as a team examples we found in participant data that both fit and did not fit. 
We also engaged in ``member checking'' with expert community partners (E9 and E10, who became participants after their establishment as community partners) throughout our data collection and analysis \cite{Saldana}. We discussed the development of our theory, particularly within the context of an online platform, and they regularly provided feedback on the feasibility and utility of the framework from their respective expertise areas. In reporting our findings, we find it important to note that while we analyzed all transcripts, we only quoted information from participants who expressed comfort in being quoted. Some participants could not be quoted due to privacy concerns related to their occupations.


\section{A Theoretical Framework of Severity}

Our goal was not to understand how specific instances in the card sorting activity were differently ranked; rather, we present a framework that describes the thinking behind severity rankings. In organizing our findings, we describe our framework of severity in two parts. Though our focus was on understanding the factors that influence how severe a particular harm might be, participants inevitably discussed these factors in terms of concrete harms. Therefore, we first present a taxonomy of four \emph{Types of Harm} that participants considered in evaluating severity: physical harm, emotional harm, relational harm, and financial harm.
We then explain the contextual factors we found that impacted severity of harm, which we call \textit{Dimensions of Severity}. These dimensions shaped participants' perspectives on what makes one harm more severe than another, and thus provide insight into how degree of harm can be assessed. 


\subsection{Types of Harm}

Participants defined the harm caused by online content, behaviors, and interactions in a multitude of ways. We start by detailing four general categories of Types of Harm that emerged from participants' discussions of harm severity, but note that they often described how these harms intertwined, informing and compounding one another. We provide broad definitions for what kinds of harm participants felt violating content might result in---a classification schema otherwise absent in current literature on online harms.


\subsubsection{Physical Harm}

\emph{Physical harm} is bodily injury to an individual or group of individuals, including self-injury, sexual abuse, or death. Though the factors that impact severity will be detailed in Section \ref{dimensions}, it is worth noting that most participants spoke about physical harm (e.g., murder) as having the highest capacity for severe harm:

\begin{quote}
    \emph{``I would put physical harm above emotional harm ... because it is most measurable and proximate, but I would - emotional harm would be a strong second.''} ---E10
\end{quote}

Physical harm could intersect with other Types of Harm. For example, physical harms are commonly accompanied by emotional harms, such as those caused by physical abuse. Furthermore, these emotional harms are not only experienced by the victims of the harm but also viewers of the harm---commercial moderators have been suffering from emotional trauma as a result of reviewing content containing physical harm \cite{Roberts}. We discuss the perspectives from which the harm is experienced in more detail in section \ref{sec:perspectives}.


\subsubsection{Emotional Harm}

\emph{Emotional harm} ranges from an annoyance (at its least severe) to a stressful or traumatic emotional response (at its most severe), whether fleeting or long-lasting. Some emotional harms were viewed as having a low capacity for causing harm, such as content like spam. As pointed to by E20, spam was viewed as more ``annoying'' than anything: \emph{``It's not actively going to endanger somebody's life, it's just annoying.''} However, many forms of emotional harm were considered to have a high capacity for harm, particularly at the intersection of both emotional and physical harm. 
E24 noted that the capacity for harm of emotional content is often underestimated: 

\begin{quote}
    \emph{``I think people way underplay emotional harms, and so I'm always cautious to be like, ``The physical is distinctly more important.'' Depictions of [direct harm to children and terrorist attacks] ... are really traumatizing to witness.''} ---E24
\end{quote}

Emotional harm, broadly, also intersected greatly with other Types of Harm. As mentioned in the last section, emotional responses are often derived from physical harms. Similarly, relational harms also had financial outcomes. Losing professional opportunities, for example, was also accompanied with stress. 

\subsubsection{Relational Harm}
\label{relational}
\emph{Relational harm} is defined as damage to one’s reputation or their interpersonal, professional, or larger community relationships. 
Relational harms were often seen as a side effect of other physical and emotional harms, like sexual exploitation. Participants discussed how survivors of non-consensual sexual imagery were often demonized or stigmatized by others, including their workplaces:

\begin{quote}
    \emph{``Anything that is non-consensual, it shouldn't be out there. It's going to harm that person. It's going to harm them emotionally, it's going to harm their income, if it gets out, it's going to harm the way people perceive them, that it might not be them, the way they want to be perceived because it's something that they don't consent to doing in the first place.''} ---G4
\end{quote}

Once more, the compounding of multiple Types of Harm was prevalent in how participants discussed harms to professional and personal relationships. 

\subsubsection{Financial Harm}

\emph{Financial harm} is defined as material or financial loss, including the loss of digital assets like accounts. Financial harm covered included harm resulting from scams (e.g., phishing), stolen or hacked accounts, and bribery and blackmail. 

Once more, financial harm intersected with other Types of Harm. For example, E8 elaborated on the financial burden of emotional blackmail as well: \emph{``Financial or content sextortion can be horrible in real-world terms.''} Financial harm was also often viewed as a side effect of other Types of Harm, such as sextortion, which can encompass all Types of Harm identified in this framework. \\

We have detailed four Types of Harm, which were often viewed as intersecting or compounding by participants: physical, emotional, relational, and financial. These Types are often attributed to harmful contents, behavior, or interactions by participants (e.g., murder coordinating being seen as physical harm). While these types themselves do not have inherent bearings of severity---for example, physical harm is not necessarily worse than emotional harm, and peaceful deaths can also be less severe than bodily injuries caused by attacks, these contextual considerations shaped the severity of particular harms, which we call \emph{Dimensions of Severity}. In the next section, we outline these eight dimensions that increase or decrease the severity of a particular harm.

\subsection{Dimensions of Severity}
\label{dimensions}

\begin{figure}[htb!] 
    \centering
    \textbf{Dimensions of Severity}\par\medskip
    \includegraphics[width=\textwidth]{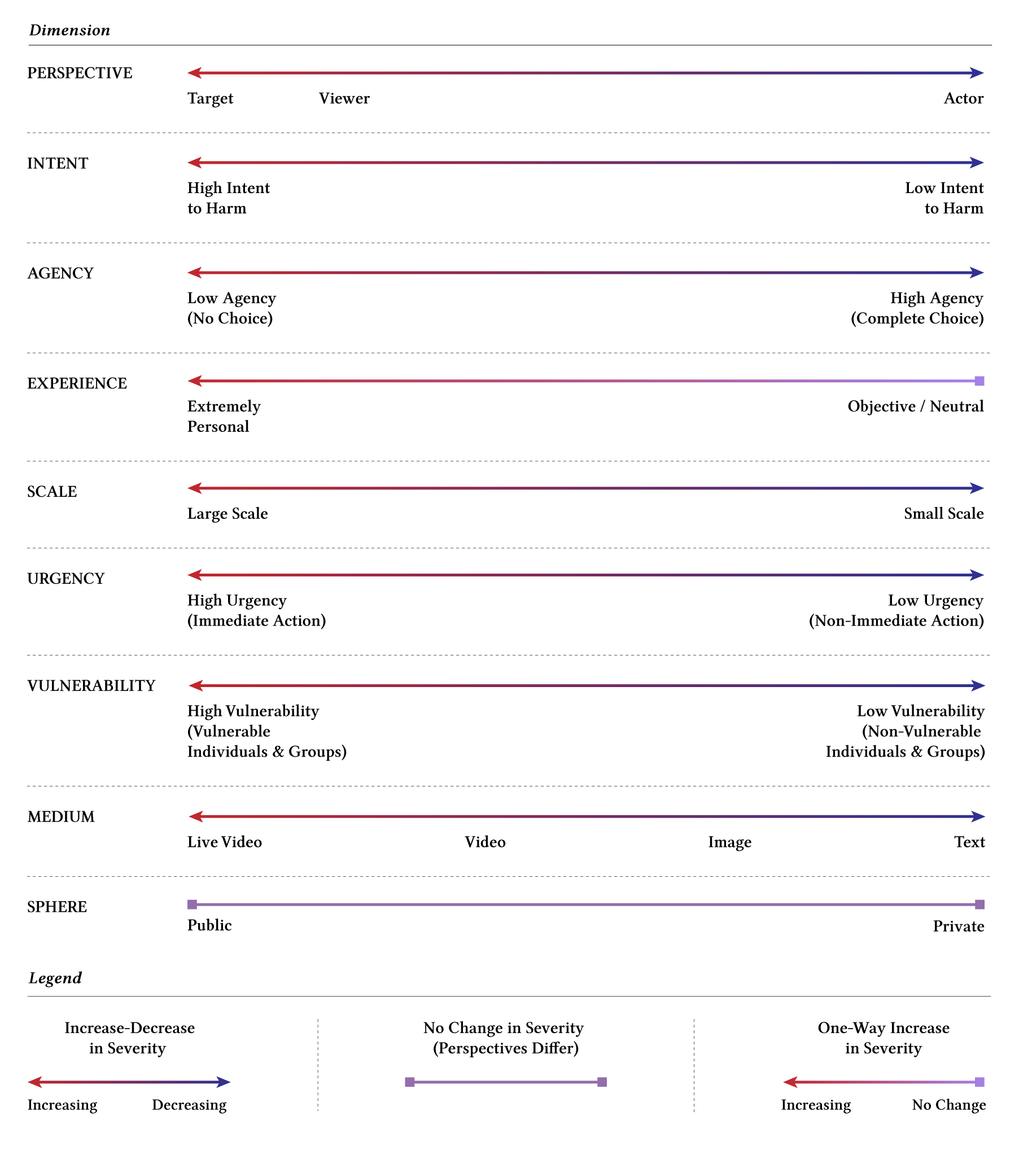}
    \caption{Dimensions of Severity on each of their associated scales. Note that some scales were different depending on the Dimension.}
    \label{fig:dimensions}
\end{figure}

In addition to the Types of Harm, participants also considered the contexts in which the harm occurred when evaluating severity. 
Through an analysis of participants' descriptions as they assessed different forms of content, we developed eight themes that shaped participant perspectives on severity. We call these themes the Dimensions of Severity, the contextual factors which either increased or decreased the level of severity of harm. The eight Dimensions we identified were: Perspectives, Intent, Agency, Experience, Scale, Urgency, Vulnerability, and Sphere. 

Each Dimension could also be more or less severe. For example, in the Dimension called Vulnerability, participants felt harm against children and animals were more severe than harm against adults. This perspective indicated that harm against individuals with high vulnerability, like that of children or animals, was more severe than harm against those with low vulnerability. These factors also influenced the way Dimensions might compound on one another to increase the severity of harm. For example, if a harm was targeted at those with high vulnerability and at a large scale (e.g., child trafficking rings that impact many children), then the combination of high Vulnerability and high Scale compounded participant perspectives on severity of the harm. Generally, participants had agreed upon understandings about increased and decreased of severity within Dimensions. The exception to this shared agreement was within the factor we called Sphere, for which opinions of whether private or public harm was worse differed even within content categories (e.g., Targeted Attacks on an Individual) (see Section \ref{Sphere}).
We demonstrate the relative scales of severity within each Dimension in Fig. \ref{fig:dimensions}.

In the following sections, we detail the definitions of Dimensions gleaned from our data, as well as how each Dimension impacted participant perspectives of severity. 

\subsubsection{Perspectives (of Actor/Viewer/Target)}
\label{sec:perspectives}
All participants discussed harm from three subject perspectives, each shaping the level of severity of the harm. First was the \textbf{\emph{actor}}, or the person contributing to the harm being caused, regardless of intent. The actor was never viewed as a victim or survivor of a harm, but rather the source of harm. Therefore, severity was attributed to their actions and the impact those actions might have on others. Further, the severity of the actor's harm on viewers and targets informed decisions about how to act against those actors. In other words, actors are not viewed as experiencing harm but causing it, but they are still relevant stakeholders when considering moderation action. For example, E1, who works on actor-level analyses of harmful content, discussed the forms of harm caused by actors, and how actors are reprimanded by what her team views as severity of those harms:

\begin{quote}
    \emph{``If it's less severe but more frequent, then the actor will probably ... not get disabled, but have a warning or a strike against that account, versus if it's very severe ... Even if someone was child grooming one time, that account should be disabled.''} ---E1
\end{quote}

The second perspective is the \textbf{\emph{viewer}}, a witness to harmful content or interactions. Viewers were often characterized as suffering most from emotional harm, often from witnessing undesirable content. The severity of harm attributed to viewers was dependent on how traumatic that content was determined to be.\emph{``I think that it would be incredibly traumatic and more harmful to just randomly come across an image or child abuse or mutilated humans.''} 

Emotional trauma might also be compounded depending on the medium of the content. For example, E3, who worked with professional content moderators, discussed how live streaming violence or sexual abuse was the most traumatic, and thus most severe, for viewers. For example, E18 imagined what it would feel like to stumble on visually graphic content: 

\begin{quote}
    \emph{``In my experience so far in this position, the worst, or most intense types of content have typically been images or videos. And I would say worst is live videos. Live streaming and that is of an individual who is defenseless and is being abused.''} ---E3
\end{quote}

The third perspective is the \textbf{\emph{target}}, the person who is, or people who are, being directly targeted by an actor. Most often, the target was considered to experience the most severe levels of harms, as the target was most associated with what participants often referred to as ``real world outcomes''---like sexual abuse, physical harm, and death. G3 earnestly expressed how they weighted the severity of harm between viewer and target: \emph{``Like what would you imagine the worst thing to happen to someone you love is. So I guess that's kind of imagining both the impact on them but also like how you're dealing with it. But I think maybe a little more focus on the impact to them.''}

During interviews and card sorting, participants often switched perspectives, particularly between targets and viewers. The roles of actors, targets, and viewers can be undulating and unstable. It is also possible that viewers or targets can become actors given a certain situation (e.g., the target of harassment starts calling the other harasser slurs). Thus, we present examples encapsulating multiple perspectives throughout the remaining Dimensions.

\subsubsection{Intent (as Perceived by a Target/Viewer)}

When participants discussed severity in relation to the actor, they did so through the lens of perceived intent ($n=31$). Intent was generally discussed when imagining whether harmful content or interactions were purposeful or not. In some cases, participants felt that the impact was more important than the intent ($n=13$)---particularly for high severity outcomes, like physical harm. In other cases, if the intent was imagined to be benign, participants felt it was more important than impact ($n=18$)---particularly for low severity outcomes, like posting images containing nudity. Participants' views on intent and its importance varied based on how severe they felt the harm was. In discussing non-sexual child nudity, G5 felt that the intent of the people posting the images---often parents or family---should be taken into account:

\begin{quote}
    \emph{``I guess I put it pretty a little down because generally pictures like that are not, personally I put a lot of store behind intent. They're typically posted by parents who think it's a cute picture and they want to put it online.''} ---G5
\end{quote}

G5 found that the outcome of child nudity online could be severe for the child. G12, also discussing child nudity, felt the opposite. They felt that, while the intent may be benign, it is the impact on the child that should be considered when thinking about the level of harm:

\begin{quote}
    \emph{``I don't think they realize that when you put something on the internet like that, it's there on the internet and regardless of what your intent behind the image was, even if it was harmless. Somebody out there is going to probably find that image or come across that image and their intent is not going to be harmless and you never know what is going to spiral out from there.''} ---G12
\end{quote}

Many participants shared G12's perspective: impact is more important than intent when assessing the severity of content, particularly for content that causes emotional and physical harm. Given the nature of intent to be difficult to solidly pinpoint, it was discussed through the perception of viewers and targets---which is also relevant to how moderators, as outside parties, would have to consider intent.

\subsubsection{Agency (of the Target and/or Viewer)} 
\label{agency}

The agency of the person harmed---whether they had a choice to participate in either the harm or circumstances leading up to the harm---was influential in how participants described ranking their decisions ($n=34$). For example, participants often felt if the person engaged in harmful behavior of their own volition, then the harm was less severe. In this case, E13 discusses how the category ``Hard Drug Sales'' was relatively mild in severity:

\begin{quote}
    \emph{``So for like drugs I mean, yes, selling drugs is illegal but at the same time you are consuming at your own risk. So, yes, that's on the bottom of my list. It's whoever presumably buying it knows what they are doing. Nobody ever buy illegal drug unknowingly that, 'Oh, I didn't know that I am buying some kind of drug.'''} ---E13
\end{quote}

On the other hand, harms associated with a lack of choice were considered more severe. This included content like mass murder coordination, bullying, suicide promotion, and sexual exploitation. 

Yet, agency was often an area of controversy, as a handful of participants actually considered some sexual abuses less severe than the majority of participants, such as non-consensual sexual imagery. For example, P11 would blame himself for sharing that imagery in the first place: 

\begin{quote}
    \emph{``Because I want to share my private pictures with this person, who I really thought this person would be cute or something like that, and then it ended up not being the case, it's just like, it will end up biting you back in the butt. That's why I kind of didn't really see it as important I guess.''} ---G13
\end{quote}

Similarly, E11 felt that, while severe, content that encourages self-harm was not as severe as non-consensual harm, because those who self-harm have to make the choice to do so:

\begin{quote}
    \emph{``A user, ultimately, may not... At the end of the day, is somebody wants to harm themselves, they didn't need [an online platform] to do that, right? } ---E11
\end{quote}

Participants' disparate feelings about certain content highlighted the underlying beliefs, cultures, and attitudes participants brought with them when considering the severity of content. In the next section, we discuss instances where participants felt one's own life experiences shaped their beliefs about severity.

\subsubsection{Experience (when Assessing a Harm)} 

Some participants elucidated that specific experiences within their own life shaped their views on the content we discussed ($n=22$). Bad experiences with certain categories themselves led them to consider them highly severe. G9 summed up this perspective in discussing the importance of considering people's personal experiences: 

\begin{quote}
    \emph{``The more personal I think it is the more severe it is. You can wage a personal attack on someone and it's more likely, when you speak personally about someone, you try to denigrate them personally over their beliefs or whatever they do, it's going to be significant more harmful.''} ---G9
\end{quote}

The subjective nature of harm was true for many participants, who openly discussed difficult life events that shaped their perspectives on severity. For example, G1 earnestly discussed her experience with rape, saying ``to me it's the worst thing that can happen.'' G4, a member of many mothers groups for addicts, felt vastly different from E14 above. Her experience led her to rank hard drug sales as highly severe:

\begin{quote}
    \emph{``I see a lot of other mothers on there that because of the sales of these type of drugs have changed their whole life from day-to-day. Their life is just hell now and it shouldn't be allowed to be something that is sold or promoted in any sort of way.''} ---G4
\end{quote}

Some participants experiences led them to make vastly different decisions than others. For example, G2 was the only participant to rank pornography as highly severe; she ranked it as the single most severe form of content. Previously a deputy sheriff, she associated pornography with violence: \emph{``Violence always comes after pornography.''} 

However, while participants acknowledged that harm was subjective, they all still tried to imagine an ``objective'' view outside their personal experiences when talking about harm---particularly harms not directly relevant to their own experiences, as highlighted in this section. Many of them tried to remove their own feelings about specific content---or they simply had little experience with it. Remaining objective in their assessment was particularly salient for expert participants discussing severity in industry contexts. In the next section, we discuss participants' conceptualizations of a dimension of severity many imagined as objective: scale. 

\subsubsection{Scale (of the Harm)} 

Participants often brought up the concept of scale---which referred to either the number of people impacted at once by the same content or event, or the amount of actors dedicated to harming an individual or group ($n=26$). The larger the scale---in terms of the number of people impacted or the number of attacks aimed at a person---the more severe many participants believed that the content was. For example, G12, who imagined murder as extremely severe, felt that the number of people murdered would increase that crime's severity further:

\begin{quote}
    \emph{``Okay. I mean, the obvious answer is to just be like a little more people are affected. I guess that's because in my eyes, murder is murder. There's nothing really worse than murder except more murder.''} ---G12
\end{quote}

Scale extended beyond individual impacts as well. Participants also felt harms that impacted society, rather than individuals, were more severe. For example, content violations like political attacks, were a greater threat to society broadly than unpleasant interpersonal interactions:

\begin{quote}
    \emph{I think political attacks on an individual level are relatively uninfluential. But I think on a broader level they've had a pretty huge impact when coordinated, though. I'm putting that high because it affects society on a significant scale.} ---E21
\end{quote}

Similarly, some participants felt that hate speech was more severe than bullying or harassment because hate speech was aimed at larger vulnerable groups and bullying and harassment tended to happen on an interpersonal level. For example, G7 imagined bullying as centralized or targeted, but hate speech as inciting larger swaths of violence against minorities:

\begin{quote}
    \emph{``Bullying is very centralized. Bullying, it deals with one particular person, okay? And it may not necessarily have anything to do with their race or the color or their nationality. It could just be a character issue with them that sets someone off to bully them.''} ---G7
\end{quote}

Other participants imagined scenarios where bullying could be more intense, when it was no longer a one-on-one interaction between the bully and the target, but a coordinated harassment campaign of many people towards one individual. The scale of harassment aimed at one individual increased the severity of the bullying category. E21 felt that a barrage of harassing content was ``very severe'':

\begin{quote}
    \emph{``So content in isolation is, I would say, less harmful ... a flood of content like a raid or an organized harassment campaign, which might include just a single piece of content from each user, is also very severe.''} ---E21
\end{quote}

Overall, participants' feelings about the severity of large scale harms led them to rank categories like ``Coordinated Attacks that Directly Harm'' people as some of the most severe. On the other hand, the prevalence of harm---scale in terms of how often it happened or was posted on platforms---did not impact the level of severity. Content with low prevalence, like child exploitation and terrorist coordination, was still more severe than content with high prevalence, like spam.

\subsubsection{Urgency (to Address the Harm)} 

Time-sensitivity impacted how participants thought of severity as well ($n=22$). The level of urgency at which participants expected an escalated emergency response correlated with how severe they felt the harm was. E11, a policymaker for the company she worked for, discussed how she prioritized the urgency in mitigating the most harm possible when creating content policies:

\begin{quote}
    \emph{``I think about it [as] how quickly do we have to [take] action to remove the content to minimize or mitigate real-world harm. So [child pornography] is a big one. We need to move in external partners. There's a lot of coordination that needs to happen. So our detection and the turnaround time for those kind of abuses is very tight.''} ---E11
\end{quote}

Meanwhile, participants felt that content that was not urgent was not severe, even for content moderation practices like content removal. E24 felt that child nudity was not important for content moderators to deal with:

\begin{quote}
    \emph{``I can't think of any strong cases where child nudity, not in the [child pornography] category, feels urgent.''} ---E24
\end{quote}

Although child pornography and child nudity both involve children, the traumatizing and exploitative nature concerned participants, making it something that (1) had to be urgently forwarded to law enforcement, for the sake of the target, the child; and (2) had to be urgently removed from social media so as not to emotionally harm viewers. Children were of particular concern to many participants. In the next section, we discuss the main concern participants had about content involving children: their vulnerability to harm.

\subsubsection{Vulnerability (of the Target and/or Viewer)} 

Vulnerability described the risk that certain people or groups of people had to being harmed. Participants described severity as higher if the target is vulnerable ($n=22$). Vulnerable groups included children and animals, as well as those determined to have less privilege or recognition in society---people of color, LGBTQ people, etc. While vulnerability is distinct from agency, those determined to have low or no agency were often considered especially vulnerable, and thus any harm towards them was most severe. Children and animals were viewed as having particularly high vulnerability and so violence and abuse of these two groups were considered egregious. 

\begin{quote}
    \emph{``Animals have no voice or any choice in what goes on with them. And they rely on humans to look after them, you know? And when we do things to intentionally or deliberately harm or kill animals, we're betraying, I would think almost a trust of nature that we've been given.''} ---G7
\end{quote}

Overall, most participants ranked physical and emotionally traumatic harm against humans as more severe than animals. The vulnerability of children meant that Direct Harm to Children as a category was consistently ranked the most severe of all the content we discussed with participants, with very few exceptions. E16 sums up her feelings on why Direct Harm to Children was ranked number one (highest in severity) for her:

\begin{quote}
    \emph{``I just put that way up there because I feel children have no way of protecting themselves, unlike adults. So I feel that is way up top.''} ---E16
\end{quote}

The vulnerability of children was considered so important to severity that most participants considered Direct Harm to Children aimed at individual children more severe than coordinated attacks against many adults. However, participants also acknowledged that certain groups of people were more vulnerable to harm than others. Hate speech was, in particular, discussed as inciting violence against historically marginalized groups. Six expert participants brought up how social media was used to mobilize against minority groups in Myanmar:

\begin{quote}
    \emph{``If you look at the top end of a hate speech and if you look at things like, Myanmar ... we totally see that like the hate of a speech resulting in real world harm by creating this humanizing devalued kind of things.''} ---E13
\end{quote}

The vulnerability of a certain individual or group impacted the relative severity of harm against that group, regardless of the content or violation. The higher participants viewed the vulnerability of a person or group, the more severe they considered harm against them.

\subsubsection{Medium (of the Harmful Content)}
\label{Medium}

Medium described the type of content that was posted---in terms of textual, visual, or audio-visual content. Every participant mentioned medium in some manner when thinking through the severity of content examples. More participants felt that visual content was worse than textual content ($n=27$). For example, G3 discussed how one could more easily stop reading descriptions of graphic violence, but seeing an image or video would leave an imprint on someone more easily:

\begin{quote}
    \emph{``If somebody were to describe graphic violence, I could just stop reading as soon as I realized the direction that was heading but when you glance at an image, you kind of get the whole thing and you can't unsee it and I think video would be the same way, right?'' ---G3}
\end{quote}

Further, livestreaming certain types of content was viewed as particularly severe, because of the knowledge that whatever abuse might be occurring was happening live. She discussed how this impacted those moderating content:

\begin{quote}
     \emph{``I would say worst is live videos ... Knowing that it's happening somewhere else in the world right now at this very moment ... `I have to take immediate action right now because what I do can affect that person right in this instant.''' ---E3}
\end{quote}

However, E5 pointed out that the severity depends on the fidelity of the content. In the case of videos, he discussed the different levels of fidelity in audio-visual content like videos and how they could be considered in terms of impacts:

\begin{quote}
    \emph{``Depending on the type of video that one is watching, like for moderators, is it in grayscale, or is it full color? Is it with volume? Is it muted? There's all sorts of nuance ... that makes a difference in the impact that it has.'' ---E5}
\end{quote}

Similarly, E13 pointed out that a piece of content may have the exact same message, but its fidelity impacts how it is handled and what its reach is. He pointed out that images and videos tend to reach larger audiences than text on social media and that ``the reach matters here a lot because the more it reaches, the more people get hurt'' (E13). However, like with other Dimensions, while there are generally aligned perceptions about the severity of certain mediums, participants acknowledged that distinctions between audio/visual/textual is not always clear:

\begin{quote}
    \emph{``Then again, it just depends, because obviously depending on what people put on, it can be very hurtful.''} ---G11
\end{quote}

Further, participants did not mention purely auditory content, such as voice messages, but prior work suggests that auditory content is often more difficult to moderate than purely textual content \cite{Jiang2019}. Given the lack of commentary on audio-based content alone, we did not include audio on Figure \ref{fig:dimensions}; understanding the relationship of audio would require future work.

\subsubsection{Sphere (the Harm Occurs In)}
\label{Sphere}
The ``sphere'' where harm took place was of consideration to some participants ($n=14$). By sphere we refer to harms that occur in the public sphere---like on public posts---versus harms that occur in the private sphere---like in private direct messages. E14 discussed how certain forms of harm often happened in the private sphere, through private messages, often making those targeted feel more alone and unsure of how to handle the situation:

\begin{quote}
    \emph{``This sextortion piece, ... usually happens in private, and people don't know who to talk to, they don't know how to reach out to anyone, they don't know what they're supposed to do about the scenario, [which] can be very dangerous in the fact that people don't feel like they have the same recourse as someone, you know, calling them a slur.''} ---E14
\end{quote}

E22, however, felt that public harassment had a ``chilling effect.'' She was cognizant of her own experiences with public harassment that made her rethink engaging with the public sphere:

\begin{quote}
    \emph{``For me and for other people I know, it means not being able to participate in public space online. It's chilling effects, there's a lot of value you can derive from these more public networks ... which me and a lot of people I know just simply cannot access because of these very real threats of just existing in public space online.''} ---E22
\end{quote}

While there was no consistent ranking of severity revealed in our analysis by the dimension of Sphere, it presented an important factor to consider when evaluating the severity of not only interpersonal harms like sextortion and harassment, but also harms for which the dimension of Sphere is not salient (such as inciting violence). For example, does a typically private harm happening in a public sphere increase its severity? Or conversely, does the typically public harm happening in a private sphere become more severe by being more concealed? We do not and cannot provide answers to these questions, and we hope that future research addresses the dimension of Sphere in more depth.

\section{Discussion}

Our Framework of Severity highlights the complexity of online harm and rejects simplistic views of them. Our work builds on critical prior work which has defined how harm is experienced in online contexts \cite{Blackwell2017a,Scheuerman2018a,Pater2016}. Specifically, we conducted this work to develop a framework of \emph{severity}, expanding on prior work to not only understand what Types of Harm occur online, but also how we might assess and reason about the relative severity when considering Dimensions of diverse harms. We did not assess differences in content-level rankings, but provide a framework for classifying, assessing, and/or thinking about \textit{what} makes some harms more severe than others. We present our framework as a tool for both researchers and practitioners examining both harm and content moderation practices.

Findings showed how contextually overlapping Dimensions of Severity can be used to understand the severity of both different Types of Harms and a wide variety of specific harmful content and interactions (e.g., graphic violence imagery). When using the taxonomy of the four Types of Harm we present---physical, emotional, relational, financial---it is important to note that they are not mutually exclusive. Some forms of harm can fall into multiple Types (e.g., sexual harassment is a complex harm that can fall into any Type of Harm). Likewise, while specific categories of harm may be more commonly associated with one Type of Harm, this is not always the case. For example, bullying is often discussed as a Type of emotional harm, but can easily become physical harm with threats of violence or injury, or resulting mental health concerns that lead to self-injury or suicide. Instead, Types of Harm should be viewed as compounding. While a harm might begin only as Financial Harm, it may compound with Physical Harm (e.g., loss of income leading to food insecurity). 

The majority of Dimensions (perspective, agency, urgency, vulnerability) colored perceptions of increased or decreased severity overall. For example, participants expressed opinions that indicated that low agency increased severity, and high agency decreased severity. While useful individually, we also found that Dimensions had the potential to compound severity. For example, the presence of high Scale and high Vulnerability might make a specific harm more severe than one that simply suffers from high-Scale. Further, the Dimension of ``experience'' had a one-way relationship with severity; content that was extremely personal for participants was perceived as more severe, but otherwise, participants attempted to remain objective or neutral in assessing harms. Our study also shows how perspectives on severity were shaped by the level of experience and expertise a participant had with the form of harm. Expert participants had different insights on the potential implications of certain harms that general population social media users might never see or interact with if not impacted personally (e.g., eating disorder promotion). In contrast with other dimensions, we found that the Dimension of ``sphere'' neither increased or decreased severity in a linear manner. Rather, participants had diverging opinions on whether public or private harms increased the severity of a harm, leading us to believe future research on public versus private contexts is necessary.

Meanwhile, we found the absence of some factors surprising. Despite the number of platform policies seemingly shaped by legal standards \cite{Fiesler2017}, we found that legality was not indicative of how our participants perceived harm---it was not a Dimension that participants considered relevant to determining severity. Interestingly, even the participants with legal expertise held the opinion that legality was not the driving factor of severity of harm. Experts working on legal policy within companies expressed this, and the academic law expert felt law was too culturally contextual for use in determining overall harm. For example, even though marijuana is moderated on platforms, and may be illegal in many countries and states in the U.S., it largely was not viewed as harmful by participants because it was not seen as embodying other dimensions which lead to negative outcomes as characterized by Types of Harm (e.g., physical harm). Similarly, child grooming and child pornography were seen as harmful not because they are illegal, but because they embodied a number of dimensions seen as leading to extremely negative outcomes.

We offer the Framework of Severity as a tool for researchers, practitioners, and policymakers to utilize in their work. In the next section, we demonstrate how the framework can be adopted for different scenarios---from granular singular harms to broad Types of Harm.



\subsection{Employing a Severity Framework in Research and Policy}

Through our analysis, we observed that participants, while reasoning through the same types of Dimensions as one another, still had differing approaches to defining severity at the content-level---often, bringing different priorities to the table, often connected to the approaches we outlined in Section \ref{domains}. Content moderation strategies also balance labor constraints \cite{Gillespie2018}, differing perspectives on governance \cite{Caplan2018}, and concerns for the emotional wellbeing of moderators \cite{Roberts}. 

Content moderation would benefit from a framework of severity when prioritizing user and platform needs. While platforms prioritize content that local regulations require they address, they might also prioritize content based on severity of harm. For example, one could imagine a platform leveraging our framework to prioritize the most severe content from the perspective of user experiences. On the other hand, a platform might use the framework to balance labor loads across more moderators to ensure moderators are not overburdened by particularly severe emotional content. Similarly, volunteer-based moderation researchers might seek to understand what kinds of content volunteer moderators regularly interact with, and which have the most severe impacts on mental wellbeing, so that they can structure training and workload to reduce trauma. As platforms make decisions about where to focus their automoderation, a severity framework can inform  efforts to alleviate the burden of moderating the most severely harmful content, allowing moderators to work on more contextual but less severe content. 

In the next few sections, we use three examples to illustrate the utility of our framework of severity by discussing how it could be used within two contexts: research and policy. Although the purpose of our framework is not to rank or quantify specific harms, we felt it would be useful to focus on harms that our participants tended to either agree or disagree on in terms of ranked severity, to demonstrate differing approaches to prioritizing disagreements on severity.

\subsubsection{Example One: Assessing Differential Severity within ``Harassment''}

The complexity of harassment is well illustrated by the GamerGate example at the beginning of this paper. Our framework allows us to consider how the harms experienced by those targeted in GamerGate is not necessarily experienced as one single umbrella form of harm, but may encompass every Type of Harm, including physical, emotional, relational, and financial \cite{Quinn2017}. Furthermore, a number of contextual Dimensions shape the severity of harm. As our framework showcases, the women harassed during GamerGate faced a complex intersection of Dimensions that made the harassment particularly severe---from being large in scale, due to the number of actors, to being urgent because of real-world threats of rape and death. Further, not all parties suffered the same level of harassment; the male journalist whose article is attributed with starting GamerGate faced little harassment from the same bad actors targeting women. Collapsing the reality of multiple, intersecting Types of Harm into a simplified ``harassment campaign'' would neglect the fact that the women suffered much more severe harassment than the men did \cite{Aghazadeh2018,Braithwaite2016,Massanari2020}. Our Framework of Severity reveals that harm is not as simple as harassing behaviors; the Types of Harm harassment encompasses, and the contextual Dimensions which shape that harm, are crucial to understanding how bad some harm can be.  

Leveraging our framework could enable scholars to design studies to identify where the most severe experiences of specific content types, like harassment, occur. Consider a scenario in which a researcher is focused on user experiences of harassment in online communities (which in our current study falls under the category of ``Targeted Attacks on an Individual Person''). The researcher might seek to understand how their participants classify the \textit{Types of Harms} harassment encompasses---is it always an emotional harm, or is there a physical component that sometimes accompanies online harassment? Perhaps that physical component is a risk associated with ongoing emotional harm, such as increased risk for self-injury or suicide. While literature on online harassment might often focus on emotional harm, in reality harassment can include other types as well.

Using our framework, researchers can map out existing literature, describing how, for instance, published work has accounted for harm across some Dimensions and Types more than others. In turn, our framework can help identify gaps in the literature where research may need to go deeper. In more established areas, we also see how our framework could help scholars develop more robust and holistic accounts of harm that can directly inform how to focus interventions into said harm. Notable here is physical harm, which has been largely absent from harassment research, yet can appear along with emotional harm, such as with increased risk for self-injury or suicide associated with harassment. Likewise, depending on the target of harassment (e.g., a professional gamer), reputational and fiscal harms may also be important to consider.

The researcher might then map out the differing experiences and perspectives on online harassment. Are there Dimensions that make certain types of online harassment more severe? Are there factors within these Dimensions which are particularly salient for severity (e.g., high Scale harassment)? Examining the Dimension of ``perspective'' could lend different insights on severity from targets, viewers, and even actors. Focusing on ``sphere'' could illuminate whether harassment in public and private settings are deferentially severe. And addressing the ``scale'' of harassment could illuminate how large scale attacks differ from smaller interpersonal experiences of harassing behaviors.

Connecting these questions to content moderation, if there are more and less severe cases, what can that tell us about more granular content moderation policies to mitigate the most severe cases of harassment? And if targets of harassment view all of their experiences as highly severe, what might that tell us about approaching the moderation of harassment more sensitively? In depth examinations of harassment guided by both \textit{Types of Harm} and \textit{Dimensions of Severity} can provide richer understandings of the range of harassment experiences and how they might vary in terms of severity. Better understanding harassment severity can lend to both platform policy and moderation practice in developing intervention points for different levels of severity.

\subsubsection{Example Two: Comparing General Population and Expert Perspectives on Severity Across ``Self-Injury and Eating Disorder Promotion''}
\label{genpop-vs-expert}

While we did not quantify such insights for the purposes of this study, card sorting revealed that participants sometimes applied the \textit{Dimensions of Severity} to specific content differently. Imagining a study specifically aimed at quantifying a rank order of severity across content categories, a researcher might be interested to understand why participants disagree so heavily on one specific content category, despite applying the same logic found in the Dimensions.

In the context of this study, industry expert and general population participants notably disagreed about the severity of the category ``Self-Injury and Eating Disorder Promotion.'' We noticed that differing perspectives around the severity of eating disorder and suicidal content arose from two differing perspectives: the urgency of Physical Harm associated with such content and the application of the Dimension of ``agency.'' Some participants ultimately believed users viewing content encouraging self-harm had agency to decide to act on that harm, and that the non-consensual nature of other physical harms was more urgent to address. While some expert participants also aligned with this viewpoint, general population participants on average viewed ``Self-Injury and Eating Disorder Promotion'' as lower severity than industry expert participants. Such perspectives differ greatly from prior work with participants struggling with eating disorders and self-injury themselves (e.g., \cite{Feuston2020,Andalibi2017a}), highlighting that the Dimension of ``experience'' may play a large role in understanding the perceived severity of content types among certain participant groups.

A researcher might design a study specifically aimed at understanding different perceptions of severity around content like eating disorders and self-injury across participant groups. Considering the findings of our own participants, one might ask: What is it about industry expert participants that make them more likely to view self-injury and eating disorder content as more severe than other participants? We might design a protocol aimed at \textit{comparing} participant logic around Dimensions. Perhaps the workplace experiences of industry experts give them certain insights into the impact of such promotional content? Perhaps general population participants are less likely to have personal experiences with eating disorders or self-injury with which to base the harm of promotional content? We might also analyze different deployments of the concept of ``agency'' to understand how choice is being operationalized, and how that may be prioritized over other Dimensions in this case (e.g., urgency, vulnerability, or experience).

Answers to such questions can inform future researcher on what participants might be most suited for assessing certain types of content and associated Dimensions. It is possible that certain participant groups are poorly suited for providing perspectives on the severity of harm of certain content categories, and should not be referenced in policy design or moderation standards. 

\subsubsection{Example Three: Comparing the Relative Severity using ``Types of Harm'' and ``Dimensions of Severity''}

How should one prioritize self-injury, hate speech, and child pornography in relationship to each other? We now turn our attention to the significant challenges that appear when thinking about forms of harm holistically at the scale of large online platforms. Research into specific experiences of harm (e.g., online harassment) provides invaluable insights that can inform how platforms engage with and address these harms. Yet, platforms must attend to all forms of harm that occur on their sites. In this section, we demonstrate the utility of our framework by considering the additional challenges that platforms experience when weighing and prioritizing what kinds of harm deserve the most attention.

Considering the relative severity of different Types of Harm may aid in prioritization efforts. Imagine a social media company that is growing rapidly struggling to handle an influx of dangerous and upsetting content given their moderation resources. The company seeks to understand how the severity of content on the platform can be broadly classified into Types of Harm, so that they can make resourcing decisions for the worst Types of Harm first. 

Alternately, a platform might start with Dimensions of Severity. Researchers might decide to first map Dimensions onto specific content categories and then classify those categories by Type, leading to insights into what Dimensions might be concentrated into which Type. Perhaps ``physical harm'' is found to have higher urgency, but ``emotional harm'' has higher scale. Such findings could lead platforms to manage content differently. Where high urgency content might be prioritized for escalation to local authorities, high scale emotional harm might require a larger moderation force trained in restorative techniques. 

Finally, research aimed at understanding how the Dimensions of Severity map onto Types could aid in creating a taxonomy of the relative severity of Types of Harm---do certain Types embody more severe forms of the Dimensions than others? Given the mass scale of content types on platforms, mapping Dimensions onto Types might help simplify moderation pipelines, allowing platforms to train moderators to handle physical, emotional, financial, and reputational harms differently.

While ranking and prioritizing forms of harm may be uncomfortable, it is an unavoidable reality given that resources for content moderation are finite. Prioritizing harm is important for at least three reasons: First, platforms have a limited number of human moderators and must allocate them strategically. As of 2019, Facebook only had 15,000 content moderators for the content generated by over 2.7 billion people \cite{Maggie2019}. But, despite debates around the right number of moderators, it is safe to assert that providing human-moderation for all content on a platform like Facebook or Twitter is not possible. Second, while tools are developed to aid moderation, engineering resources are also finite. Just as with human moderation, platforms must decide what Types of Harm their engineering efforts should focus on as they develop automated ways of detecting and addressing different forms of harm. Finally, even if resources to reduce harm were increased dramatically and platforms made a fantastical commitment to eliminate all harm, they would still need to prioritize forms of harm in the interim as they worked towards that eventual goal.


Deciding on the right approach to reducing harm in online communities is not a trivial undertaking, and we do not assert that there is a single correct solution. Rather we suggest that the only inappropriate approach is to not consider the severity of the harm, and we believe that our framework can aid in prioritizing efforts for a given platform or context. We believe our framework can help provide nuance to current practices that platforms use while addressing harm. A common approach when it comes to content moderation in online communities and community standards is to describe content or actions in topical groups or categories \cite{JialunAaronJiang2020}, such as the ones we used in this study. One can imagine how online communities seeking to reduce harm might use these categories to prioritize efforts. Yet, as we have seen in our work, harm is not uniform with a given category of violation. When considering the harm associated with violations (as opposed to, say, legal requirements associated with violations) using this framework to develop contextually appropriate ways of understanding the different levels of severity within a category might better inform and support efforts.

\section{Limitations and Future Work}
We acknowledge that there are limitations with both our methods and the scope of this work. In terms of methods, our sample skews largely towards specific regions---North America and South Asia. It is possible that new Types and Dimensions could be theorized by engaging with experts and general population participants from other regions. There is opportunity to expand the proposed severity framework with more perspectives from across a number of difference cultures.

Further, we did not assess content-level severity differences in this work. Therefore, we do not offer any quantitative or statistical insight into which content categories are most severe and which are least severe (e.g., proof that Direct Harm to Children is quantifiably and provably worse than another violation, like Spam). We also cannot offer insight into \textit{how much} worse certain harms might be for participants. Further, while, in the context of our framework, participants from sampled regions attended to the same Types and Dimensions of harm in their thinking about severity, content-level perspectives on severity may differ by culture---regionally or nationally. Similarly, we did not quantify differences between experts and general population social media users. Future work measuring the either the severity of categorical (e.g., Direct Harm to Children) or content-specific (e.g., child grooming) rankings would bolster attempts to understand harm online.

Finally, while our framework does not prescribe an approach to prioritizing specific harms based on severity, it offers a first step towards assessing potential prioritization frameworks in research and moderation. As demonstrated through our Discussion, thinking through research problems utilizing our severity framework can aid in developing different approaches to handling harm online. We encourage future research to employ our framework to explore methods of prioritization based on the severity of harm.

\section{Conclusion}

While research to date on specific forms of online harm has been nuanced, thorough, and survivor-centered, we offer a high-level perspective informed by both the general public and domain experts. We developed a theoretical framework of severity for online harm through empirical interviews with 52 participants. Both experts in harm and content moderation work and general population social media users provided rich data on the relationships between a variety of online harms. Our framework offers researchers and practitioners an empirically-grounded tool for which to assess the severity of singular or multiple forms of online harms for research, policy, and moderation purposes. Through use of the Framework of Severity, researchers and practitioners can adopt, critique, shift, and apply it to a variety of contexts. For example, general population studies, expert communities, and survivor and target communities. Researchers and practitioners can contextualize their work by utilizing specific aspects of our framework, while understanding they may be unable to capture every dimension of severity for their specific project. 
We demonstrated its utility by discussing the challenges to mitigating online harm, and how assessing the severity of that harm by use of our framework can help guide informed prioritization approaches. 

\begin{acks}
We'd like to thank Jess Bodford, Frank Kayanet, Mar Drouhard, Jes Feuston, Katy Weathington, Mally Dietrich, and Frank Stinar for discussions and feedback on this work. Finally, we'd like to thank Katie Gach, Katy Weathington, and Samantha Dalal for copyediting the final version of this paper. This research was partially supported by funding from Facebook.
\end{acks}

\received{October 2020}
\received[revised]{April 2021}
\received[accepted]{July 2021}

\bibliographystyle{ACM-Reference-Format}
\bibliography{.CSCW2021}

\end{document}